# Induced antiferromagnetism and large magnetoresistances in $RuSr_2(Nd,Y,Ce)_2Cu_2O_{10-\delta}$ ruthenocuprates

A. C. Mclaughlin*

Department of Chemistry, University of Aberdeen, Meston Walk, Aberdeen AB24 3UE, UK.

F. Sher

Chemical and Materials Engineering Department, Pakistan Institute of Engineering and Applied Sciences, Nilore, Islamabad, Pakistan

S. A. J. Kimber and J. P. Attfield

Centre for Science at Extreme Conditions and School of Chemistry, University of Edinburgh, King's Buildings, Mayfield Road, Edinburgh EH9 3JZ.

**Abstract**

$RuSr_2(Nd,Y,Ce)_2Cu_2O_{10-\delta}$ ruthenocuprates have been studied by neutron diffraction, magnetotransport and magnetisation measurements and the electronic phase diagram is reported. Separate Ru and Cu spin ordering transitions are observed, with spontaneous Cu antiferromagnetic order for low hole doping levels $p$, and a distinct, induced-antiferromagnetic Cu spin phase in the $0.02 < p < 0.06$ pseudogap region. This ordering gives rise to large negative magnetoresistances which vary systematically with $p$ in the $RuSr_2Nd_{1.8-x}Y_{0.2}Ce_xCu_2O_{10-\delta}$ series. A collapse of the magnetoresistance (MR) and magnetisation in the pre-superconducting region may signify the onset of superconducting fluctuations.

INTRODUCTION

In the past 20 years layered copper oxides such as $La_{2-x}Sr_xCuO_4$ [1] and $YBa_2Cu_3O_{6+x}$ [2] have been extensively studied due to the observation of high-temperature superconductivity (HTSC). A final theory of HTSC has yet to be established although there is increasing evidence that the pairing mechanism is mediated by magnetic excitations [3, 4]. Recently a sharp magnetic excitation or resonance has been observed by neutron diffraction in both hole doped [4] and electron doped materials [3]. The resonance energy $E_r$ scales with $T_c$ so that $E_r \sim 5.8k_BT_c$ for both hole and electron doped materials demonstrating that it is a fundamental property of the superconducting copper oxides. Another leading issue is the observation of a pseudogap below a characteristic temperature T* in underdoped cuprates. It is still unclear whether the pseudogap represents a new electronic state which competes with the superconducting phase or if the pseudogap state represents a precursor to the superconducting d-wave gap[5,6] but further knowledge of this state may lead to greater insight of the superconductivity mechanism. Large bulk negative magnetoresistances (MR = $((\rho_H-\rho_0)/\rho_0)$) have recently been observed in underdoped 1222 ruthenocuprates at low temperatures, up to MR = -49 % at 4 K and 9 T in $Ru_{0.8}Ta_{0.2}Sr_2Nd_{0.95}Y_{0.15}Ce_{0.9}Cu_2O_{10-\delta}$, [7, 8 9 10], demonstrating strong spin-charge coupling within the $CuO_2$ planes. Significant MR is observed over a wide hole doping range ($0.01 \leq p \leq 0.06$) [8] and hence can be used to probe the electronic states in the pseudogap region.

Layered ruthenocuprate materials $RuSr_2RECu_2O_8$ [11, 12], $RuSr_2(RE_{2-x}Ce_x)Cu_2O_{10-\delta}$ [13] (RE = Sm, Gd, Eu) and $Pb_2RuSr_2Cu_2O_8Cl$ [14] have been extensively studied due to the presence of coexisting weak ferromagnetism and superconductivity. Large −MR's are observed in the new 1222 type ruthenocuprate series $RuSr_2Nd_{1.8-x}Y_{0.2}Ce_xCu_2O_{10-\delta}$ ($0.7 \leq x \leq 0.95$) at low temperatures [7, 8]; -MR initially rises to ~2% below the Ru spin ordering temperature, $T_{Ru}$, as observed in other superconducting ruthenocuprates but increases dramatically on cooling. Neutron diffraction studies have shown that below $T_{Ru}$, additional peaks from a (½ ½ ½) magnetic superstructure are observed which can be fitted by a model of antiferromagnetically ordered Ru moments aligned in the *c* direction [7]. The Cu spins order antiferromagnetically in the *ab* plane with a (½ ½ 0) superstructure below a second transition at $T_{Cu}$, as shown in Fig. 1. The variation of –MR with temperature and field are characteristic of charge transport by magnetopolarons – small ferromagnetic regions surrounding each Cu-hole within a matrix of antiferromagnetically ordered $Cu^{2+}$ spins.[15] An applied magnetic field cants the Ru spins into a ferromagnetic arrangement, which induces partial ferromagnetism in the $CuO_2$ planes thereby increasing the mobility of the magnetopolarons, giving the observed, negative MR's [8]. Magnetopolaron hopping is a thermally activated process, leading to a characteristic exponential rise in –MR [16] below the Cu spin transition $T_{Cu}$[7].

The magnetotransport in 1222 ruthenocuprates is also very sensitive to lattice effects. In a series of $RuSr_2R_{1.1}Ce_{0.9}Cu_2O_{10-\delta}$ (R = Nd, Sm, Eu, and Gd with Y) samples where the hole doping level is constant, the high field MR does not correlate with the paramagnetic moment of the *R* cations, but shows an unprecedented crossover from negative to positive MR values as $<r_A>$, the mean A site ($R_{1.1}Ce_{0.9}$) cation radius decreases [8]. This lattice effect is further evidenced from studies of $Ru_{1-x}Ta_xSr_2Nd_{0.95}Y_{0.15}Ce_{0.9}Cu_2O_{10}$ materials; $MR_{9T}$(4 K) increases from -28% to -49% as *x* increases from $0 - 0.2$ which further expands the unit cell [10]. The latter MR is the largest reported for ceramic copper oxide materials at this field strength and is comparable to values (at higher temperatures) in spin-polarized conductors such as the CMR manganites [17, 18] and $Sr_2FeMoO_6$ [19].

In this paper, further results from a neutron diffraction, magnetoresistance and magnetic study of the ruthenocuprate series $RuSr_2Nd_{1.8-x}Y_{0.2}Ce_xCu_2O_{10-\delta}$ will be presented. The results demonstrate that the magnetic order in the $RuO_2$ planes induces long range antiferromagnetism in the $CuO_2$ plane across the pseudogap region, which results in the large magnetoresistances.

SAMPLE PREPARATION AND ANALYSIS

Samples of $RuSr_2Nd_{1.8-x}Y_{0.2}Ce_xCu_2O_{10-\delta}$ ($x$ = 0.70, 0.75, 0.80, 0.85, 0.90, 0.95) and $RuSr_2Nd_{1.0}Y_{0.1}Ce_{0.9}Cu_2O_{10-\delta}$ were prepared by the solid-state reaction of stoichiometric powders of $Nd_2O_3$, $Y_2O_3$, $RuO_2$, $CuO$, $CeO_2$ and $SrCO_3$, as described elsewhere [8]. Part of the "as prepared" $x$ = 0.90 and 0.95 samples were annealed for 24 hours under flowing $N_2$ at 600 ºC and furnace cooled to form more oxygen-deficient samples. Part of the "as prepared" $x$ = 0.70, 0.75 and 0.80, samples were annealed for 24 hours under flowing $O_2$ at 800 ºC to reduce the oxygen deficiency $\delta$.

Room temperature X-ray diffraction patterns were collected on a Bruker D8 Advance diffractometer with twin Gobel mirrors using Cu K$\alpha_1$ radiation. Data were collected over the range 5 °< 2$\theta$ < 100 °, with a step size of 0.02 °. The profiles could all be indexed on a tetragonal *I4/mmm* symmetry space group as previously reported for the 1222 ruthenocuprates[20]; a small amount of a secondary 1212 phase $RuSr_2(Nd,Y)Cu_2O_8$ is present in all samples.

The unit cell of $RuSr_2Nd_{1.8-x}Y_{0.2}Ce_xCu_2O_{10-\delta}$ consists of $RuO_2$, $SrO$ and $CuO_2$ layers are stacked as in $RuSr_2GdCu_2O_8$, but in this case the $CuO_2$ layers are separated by a $Nd_{1.8-x}Ce_xY_{0.2}O_2$ block rather than a single rare earth layer. It has previously been reported that oxygen vacancies in $RuSr_2Gd_{2-x}Ce_xCu_2O_{10-\delta}$ are located on the O(4) sites within the $Gd_{2-x}Y_{0.2}Ce_xO_2$ block and that the concentration of oxygen vacancies increases as $x$ decreases [20]. The oxygen contents of the present samples were determined by thermogravimetric analysis in a 5% $H_2/N_2$ atmosphere using a Stanton Redcroft 780 thermal analyser. Data were recorded at a rate of 3 °C/minute between temperatures of 20 - 800 °C. The samples decomposed to a mixture of Cu, Ru, SrO, $CeO_2$, $Nd_2O_3$ and $Y_2O_3$ in two steps between 300 - 750 °C, enabling $\delta$ to be determined from the mass loss. The results in Table 1 confirm that the oxygen deficiency decreases with Ce content in $RuSr_2Nd_{1.8-x}Y_{0.2}Ce_xCu_2O_{10-\delta}$; as $x$ increases from 0.7 to 0.95, $\delta$ changes from 0.0905 to 0.015.

Thermogravimetric analysis also shows that $N_2$-annealing increases $\delta$ whereas annealing the samples under flowing $O_2$ results in a decrease in $\delta$. XANES studies have shown that Ru remains in the formal +5 state in the 1222 ruthenocuprates (although this is not true of 1212 types), e.g. the measured Ru valence remains at 4.95(5) as $x$ increases from 0.5 to 1.0 in $RuSr_2Gd_{2-x}Ce_xCu_2O_{10-\delta}$[24], and so reliable Cu hole doping concentrations $p$ can be calculated from the cation and oxygen contents as $p = (1 - x - 2\delta)/2$, giving the values shown in Table 1.

## MAGNETISATIONS

Magnetisations were measured between 5 K and 300 K on a Quantum Design SQUID magnetometer in an applied field of 100 Oe after zero-field (ZFC) and field cooling (FC), and data for representative $RuSr_2Nd_{1.8-x}Y_{0.2}Ce_xCu_2O_{10-\delta}$ samples are displayed in Fig. 2. Neutron diffraction (see later) shows that two magnetic transitions occur; canted antiferromagnetic (weak ferromagnetic) Ru spin ordering at $T_{Ru}$ and antiferromagnetic order of the Cu spins at a lower temperature transition $T_{Cu}$. A deviation between ZFC and FC data is evidenced for all samples. In the ZFC data M/H initially increases due to the weak ferromagnetic ordering of the Ru spins below $T_{Ru}$ and at lower temperatures M/H is reduced as the Cu spins order antiferromagnetically. The irreversibility observed in ZFC and FC data arises due to a Dzyaloshinsky-Moriya (DM)[21, 22] exchange between neighbouring Ru sites which is possible due to the tilts and rotations of the $RuO_6$ octahedra around $c$[20]. The field causes both the Ru and Cu spins to cant out of their original direction and align a component of their moments with the direction of H[7, 23].

It has previously been shown that $T_{Ru}$ can be determined by extrapolating the maximum (-dM/dT) slope to zero magnetization, while $T_{Cu}$ is estimated from the temperature of the maximum zero field cooled magnetization ($M_{max}$)[7, 8]. These parameters are shown in Table I and the transition temperatures have been used to construct the phase diagram in Figure 3. Both $T_{Ru}$ and $T_{Cu}$ decrease as the Cu hole doping level p increases through changes in x and $\delta$ in our RuSr$_2$Nd$_{1.8-x}$Y$_{0.2}$Ce$_x$Cu$_2$O$_{10-\delta}$ samples. A similar decrease in $T_{Ru}$, from 110 to 50 K, is seen as x decreases from 1.0 to 0.5 in RuSr$_2$Gd$_{2-x}$Ce$_x$Cu$_2$O$_{10-\delta}$ [24]. Furthermore, insertion of hydrogen into RuSr$_2$Eu$_{1.4}$Ce$_{0.6}$Cu$_2$O$_{10-\delta}$, which reduces the hole content, increases $T_{Ru}$ from 92 K to 167 K. [25] Although $T_{Ru}$ decreases monotonically with doping in RuSr$_2$Nd$_{1.8-x}$Y$_{0.2}$Ce$_x$Cu$_2$O$_{10-\delta}$ (Fig. 2) $T_{Cu}$ shows a discontinuity near p = 0.02. This is the doping limit for long range antiferromagnetism in simple cuprates, and neutron diffraction shows that this is the boundary between normal (spontaneous) and induced antiferromagnetic Cu spin phases (see later).

## MAGNETOTRANSPORT MEASUREMENTS

Magnetoresistance measurements were performed on sintered polycrystalline bars (approximate dimensions 4 x 4 x 12 mm$^3$) between 4 and 290 K using a standard four–probe ac technique on a Quantum Design Physical Property Measurement System in magnetic fields up to 7 T. All of the RuSr$_2$Nd$_{1.8-x}$Y$_{0.2}$Ce$_x$Cu$_2$O$_{10-\delta}$ samples are semiconducting down to 4 K and do not show superconducting transitions or resistive anomalies at the magnetic transitions; representative plots are shown in Fig. 4.

The variation of MR with H for the RuSr$_2$Nd$_{1.8-x}$Y$_{0.2}$Ce$_x$Cu$_2$O$_{10-\delta}$ samples is displayed in Figure 5. Large negative magnetoresistances are observed at 5 K for all samples. Microstructural effects such as domain or grain boundary resistances are evident in the low field responses, as the MR-H curvature is negative for as–prepared samples but positive for O$_2$ and N$_2$-annealed materials. An increase in grain boundary resistances may also account for the anomalously high resistivity of the N$_2$-annealed x = 0.9 sample (Fig. 4). However, the high field MR's are linear and the high field values (Table 1 and Fig. 6) show a strikingly smooth variation with the hole-doping level *p*, suggesting that intrinsic factors dominate[8]. The initial rise of -MR$_{7T}$(5K) for *p* = 0 to 0.04 has been attributed to the increase in the number of holes acting as magnetopolaron carriers in the $CuO_2$ planes. The applied magnetic field cants the Ru and Cu spins towards a parallel alignment increasing the magnetopolaron mobility and resulting in large -MR. The peak in $-MR_{7T}(5K)$ at $p = 0.04$ does not correspond to any discontinuity in $T_{Ru}$ and $T_{Cu}$ and so does not indicate a phase boundary, but is accompanied by the onset of a sharp decline of the maximum magnetisation $M_{max}$ at $T_{Cu}$ above $p = 0.033$, also shown in Fig. 6. This suggests that magnetopolarons in the $CuO_2$ planes are lost above $p = 0.04$ due to the onset of d-wave pairing correlations between the Cu holes, eventually leading to superconductivity for $p > 0.06$. Thus the magnetoresistance is sensitive to the competition between antiferromagnetic and superconducting fluctuations in the induced (i-AF) phase.

## NEUTRON DIFFRACTION

### A. Structure Refinements

Variable temperature neutron diffraction patterns of $RuSr_2Nd_{1.8-x}Y_{0.2}Ce_xCu_2O_{10}$ (x = 0.80, 0.90, 0.95) and $RuSr_2Nd_{1.0}Y_{0.1}Ce_{0.9}Cu_2O_{10}$ for structure refinements were recorded at a wavelength of 1.5943 Å on instrument SuperD2B at the ILL, Grenoble. A 5g sample was inserted into an 8mm vanadium can and neutron diffraction patterns were recorded with an acquisition time of one hour at 5, 30, 60, 90, 120, 150, 190, 220 and 290 K. Time of flight neutron diffraction data were recorded for the $N_2$-annealed sample of $RuSr_2Nd_{0.9}Y_{0.2}Ce_{0.9}Cu_2O_{10-\delta}$ on the high intensity instrument POLARIS at the ISIS Facility, Rutherford Appleton Laboratory. A 5 g sample was inserted into an 11mm vanadium can and neutron diffraction patterns were recorded with an acquisition time of one hour at 5, 30, 60, 100 and 140 K, (but data collected at higher temperatures were unusable due to a technical problem).

The neutron diffraction patterns were all fitted by the Rietveld method [26] using the GSAS program [27]. The backgrounds were fitted using linear interpolation and the peak shapes were modelled using a pseudo–Voigt function. Excellent Rietveld fits were obtained (Figure 7) for all profiles using a tetragonal *I4/mmm* structural model. There was no evidence of superstructure or orthorhombic distortion, but disordered rotations and tilts of the $RuO_6$ octahedra were evidenced as previously seen in $RuSr_2GdCu_2O_8$ [12], $Pb_2RuSr_2Cu_2O_8Cl$ [14] and $RuSr_2Gd_{1.5}Ce_{0.5}Cu_2O_{10-\delta}$ [20]. These were modelled by splitting the oxygen sites of the $RuO_6$ octahedra. All of the metal occupancies refined to within ± 1% of full occupancy and there was no evidence for cation anti-site disorder. Variable temperature results for the x = 0.95 sample are shown in Table II, and similar tables for the other samples have been deposited as auxiliary material [28]. Refined cell parameters and selected bond lengths at 5 K for all the samples studied are displayed in Table III.

The results summarised in Table III show that the 1222 structure does not change smoothly as a function of hole doping alone, but also varies with the Nd/Y/Ce and oxygen contents. Comparison of the three samples with Ce content x = 0.9 shows that both an increasing oxygen deficiency $\delta$ (in the $N_2$-annealed sample) which removes holes, and the size effect of replacing some $Y^{3+}$ by larger $Nd^{3+}$ (compare $N_2$-annealed and y = 0.1 samples, which have approximately constant $\delta$) expand both the lattice parameters. Within the as prepared $RuSr_2Nd_{1.8-x}Y_{0.2}Ce_xCu_2O_{10-\delta}$ series, as x decreases from 0.95 to 0.80, the size effect of replacing $Ce^{4+}$ with larger $Nd^{3+}$ counteracts the volume reduction due to increased hole doping, and no clear trend in *a* or cell volume emerges, although *c* contracts across the three samples. A small shift of the R cations away from the $CuO_2$ planes as the oxygen deficiency $\delta$ increases is evidenced by the variation of the R-O(2) bond in Fig. 8. This is likely to a consequence of stronger bonding between R cations and the remaining O(4) atoms as vacancies are created in this layer.

The lattice parameters are plotted as a function of temperature in Figure 9. A typical Debye-like thermal expansion of the lattice is observed for all samples and there is no evidence for the negative thermal expansion below $T_{Ru}$ that was erroneously reported before[7]. Although no magnetostriction is seen in the lattice parameters, an anomaly is apparent when the contributions of the interplanar Cu-Cu distance $d_I$ and the thickness of the ruthenocuprate slab $d_{RC}$ to the c-axis length (Fig. 1) are considered. Fig 10 shows a clear change in slope of $d_I$ at $T_{Ru}$ for three of the four samples, and in two cases $d_I$ expands on cooling below $T_{Ru}$. By contrast, no anomalies are observed in the thermal variation of $d_{RC}$. A similar anomaly has previously been observed at the Ru spin ordering temperature from neutron diffraction measurements on $RuSr_2GdCu_2O_8$ [29]. The change of slope for $d_I$ does not correlate with the hole doping p, but does follow the oxygen deficiency $\delta$, with the largest anomaly for the smallest $\delta$ (Figs 8 and 10). The origin of this effect is unclear. The oxygen vacancies may frustrate the magnetic exchange interactions between $RuO_2$ layers, weakening the magnetostriction.

## B Magnetic Diffraction

To observe weak magnetic diffraction peaks, variable temperature neutron diffraction patterns were recorded for $RuSr_2Nd_{1.0}Y_{0.1}Ce_{0.9}Cu_2O_{10-\delta}$ and $RuSr_2Nd_{0.9}Y_{0.2}Ce_{0.9}Cu_2O_{10-\delta}$ on the high intensity instrument D20 at the ILL at a wavelength of 2.4189 Å. A 5g sample was inserted into an 8mm vanadium can and the temperature was increased from 5 to 290 K at a rate of 1 K per minute. Neutron diffraction data were recorded continuously and summed every 5 K. Neutron diffraction patterns were also recorded for $RuSr_2Nd_{1.0}Y_{0.2}Ce_{0.8}Cu_2O_{10-\delta}$ on D20 at temperatures 4, 10 , 24, 40 , 60  80 , 100  and 120 K.

We previously reported that the Ru spins order with a (½ ½ ½) propagation vector and moments aligned parallel to the *c* direction in $RuSr_2Nd_{0.9}Y_{0.2}Ce_{0.9}Cu_2O_{10-\delta}$ below $T_{Ru} = 130$ K [7]. As the temperature is decreased further the Cu spins order antiferromagnetically in the *a-b* plane with a (½ ½ 0) superstructure (Fig 1). The temperature variation of the magnetic superstructure intensities for this (p = 0.033) phase and for $RuSr_2Nd_{1.0}Y_{0.2}Ce_{0.8}Cu_2O_{10-\delta}$ (p = 0.055) are compared to those for the less doped (p = 0.017) $RuSr_2Nd_{1.0}Y_{0.1}Ce_{0.9}Cu_2O_{10-\delta}$ in Figure 11. The overlapped [(½ ½ ½) + (½ ½ 0)] intensity and the (½ ½ 2) peak intensity are sensitive to the magnitudes of the ordered Ru and Cu moments, respectively. The thermal evolutions of the [(½ ½ ½) + (½ ½ 0)] intensities show that the Ru spin orderings are spontaneous in all three materials, with a sharp magnetic transition below which the intensities rapidly rise to a saturation value (taken to be the 4 K intensity, $I_{4K}$). These data can be described by the critical expression $I/I_{4K} = (1-T/T_c)^{2\beta}$; and fits in the $T_c/2 < T < T_c$ regions give $\beta = 0.35$, 0.26 and 0.25 for p = 0.017, 0.033 and 0.055 samples, respectively. The value for the p = 0.017 sample is typical of three-dimensionally ordered antiferromagnets, however the lower values for the p = 0.033 and 0.055 samples suggest a more two-dimensional character. Two distinct behaviours for Cu spin order are revealed by the thermal variations of I(½ ½ 2). Spontaneous Cu antiferromagnetism is observed for the p = 0.017 sample, and the critical fit gives $\beta = 0.38$, close to the value for Ru order.

The thermal evolution of the (½ ½ 2) intensity for both the p = 0.033 and p = 0.055 samples (Fig. 11) are different to that for p = 0.017, as the intensity rises gradually below an ill-defined transition, and is not saturated down to 4 K. This is characteristic of an induced (non-spontaneous) magnetic order. The difference between the Cu spin orderings at doping levels of p = 0.017 and p = 0.033 is consistent with the discontinuity in $T_{Cu}$ from magnetization data seen at p = 0.021 on Fig. 3, and confirms that this is the boundary between spontaneous and induced antiferromagnetic phases, with long range (½ ½ 0) Cu spin order in both cases. This demonstrates that the Ru spin order induces Cu spin order across the $0.02 \leq p \leq 0.06$ pseudogap region where long range Cu antiferromagnetism is not normally observed in cuprates.

## DISCUSSION

The 1222 type ruthenocuprates are difficult materials to prepare. Single crystals or epitaxial films have not been reported, and polycrystalline samples always contain traces of 1212 type or other secondary phases. The intrinsic chemistry of 1222 phases is also complex. The present study of $RuSr_2Nd_{1.8-x}Y_{0.2}Ce_xCu_2O_{10-\delta}$ demonstrates that both the $R^{3+}/Ce^{4+}$ ratio and the deficiency of oxygen in the $(R,Ce)_2O_{2-\delta}$ block are important in controlling the hole doping p of the $CuO_2$ planes. The clear trends of magnetic transition temperatures (Fig. 3) and magnetoresistances (Fig. 6) with the experimentally determined p values shows that these are meaningful, and that it is important to take oxygen deficiency into account when considering physical trends in 1222 materials.

Neutron diffraction (Fig. 11) shows that both the Ru and Cu spins order at distinct transitions. While the variable temperature neutron experiments provide direct measurements of $T_{Ru}$ and $T_{Cu}$, useful estimates of the transition temperatures can be derived from the low field magnetisation measurements (Fig. 2), at least for comparative purposes. By combining these values with those of a superconducting $RuSr_2Nd_{1.4}Ce_{0.6}Cu_2O_{10-\delta}$ sample [30] and the $RuSr_2Gd_{2-x}Ce_xCu_2O_{10-\delta}$ system [31] which overlaps the superconducting regime, the composite electronic phase diagram for ruthenocuprates shown in Fig. 3 is constructed. $T_{Ru}$ decreases with increased hole-doping of the $CuO_2$ planes, but the discontinuity between the Nd- and Gd-based series at p = 0.06 shows that R-cation size effects are also important. The magnetic coupling between $RuO_2$ layers becomes stronger as the thickness of the $(R,Ce)_2O_{2-\delta}$ block (Fig. 1) is decreased on changing from larger $Nd^{3+}$ to smaller $Gd^{3+}$. The importance of Ru-Ru coupling through the $(R,Ce)_2O_{2-\delta}$ block is corroborated by the structural anomaly in $d_l$ at $T_{Ru}$ (Fig.10) which is sensitive to the oxygen deficiency $\delta$. Furthermore $RuSr_2Nd_{1.8-x}Y_{0.2}Ce_xCu_2O_{10-\delta}$ samples with x = 0.8 ($O_2$ annealed) and x = 0.7 (as-prepared) which have the same hole concentration, p = 0.055, but distinct values of $\delta$ (Table 1) exhibit significant differences in the magnetic parameters $T_{Ru}$, $T_{Cu}$ and $M_{max}$ so that $T_{Ru}$ decreases from 90 K to 78 K as $\delta$ increases from 0.045 to 0.095 (Table 1). Hence the magnetic transition temperatures $T_{Ru}$ and $T_{Cu}$ are dependent both on the hole concentration, p and on the oxygen deficiency in the $(R,Ce)_2O_{2-\delta}$ block.

Fig. 3 shows that coupling between the Ru and Cu spins in the low doped ruthenocuprates extends the limit of long range antiferromagnetic Cu spin order up to the border of superconductivity at 6% hole doping. The p = 0.02 discontinuity in $T_{Cu}$ marks the boundary between the spontaneous antiferromagnetic phase that is generic to cuprates, and an induced-Cu antiferromagnetic phase in the $0.02 < p < 0.06$ region where a 'mixed' phase without long range magnetism is usually observed. This shows that the near-antiferromagnetic Ru spin order aligns the locally correlated Cu antiferromagnetic regions intrinsic to the mixed phase. We deduce that a direct boundary between induced-Cu antiferromagnetism and superconducting phases is present near $p \approx 0.06$. It will be useful to prepare further 1222 series that span this boundary to study directly the low temperature antiferromagnetism to superconducting transition. The induced nature of the Cu antiferromagnetism from $0.02 < p < 0.06$ is further corroborated by variable temperature neutron diffraction measurements on a sample in the AF region (p = 0.017) and two samples in the i-AF region of the phase diagram with p = 0.033 and p = 0.055 (Fig 11). The thermal evolution of the (½ ½ 2) magnetic intensity for samples with $p < 0.02$ can be described by a critical expression. In contrast the intensity of the (½ ½ 2) peak for samples with p>0.02 increases gradually with decreasing temperature and there is no evidence of saturation down to the lowest temperature recorded which is characteristic of induced magnetic order.

Structural studies of $RuSr_2Nd_{1.8-x}Y_{0.2}Ce_xCu_2O_{10-\delta}$ by neutron diffraction show that a magnetostriction in the interplanar $CuO_2$ separation, $d_I$ is observed at $T_{Ru}$ so that there is a clear change in slope of $d_I$ at $T_{Ru}$ for three of the four samples (Figures 8 and 10). A similar anomaly has previously been observed at the Ru spin ordering temperature from neutron diffraction measurements on superconducting $RuSr_2GdCu_2O_8$ [29]; the origin of this magnetostriction is not yet known. However it has been demonstrated that the Ru spin ordering induces antiferromagnetic order in the $CuO_2$ plane from p = 0.02 – 0.059 and hence the magnetostriction in $d_I$ could be a consequence of induced magnetic clusters within the $CuO_2$ plane at $T_{Ru}$ which grow in size with decreasing temperature so that antiferromagnetic order is detected by neutron diffraction at lower temperatures. μSR experiments are underway to investigate this further.

Large negative magnetoresistances have previously been observed in lightly doped cuprates [32, 33, 34] as Cu spin reorientation of the antiferromagnetic phase in an applied field enhances the conductivity, particularly the c-axis transport. Results show that there is no significant variation in the MR upon increasing the oxygen deficiency, $\delta$ (Table 1). This is clearly demonstrated by the magnetotransport properties of samples with x = 0.8 ($O_2$ annealed) and x = 0.7 (as-prepared) which have the same hole concentration (p = 0.055) but significant differences in oxygen deficiency; $MR_{7T}$ (5 K) = -19 % for both samples. Ru spin order in the ruthenocuprates extends Cu antiferromagnetism to higher dopings, enabling the trend of high-field MR with p (Fig. 6) to be discovered. As p increases to 0.042 in $RuSr_2Nd_{1.8-x}Y_{0.2}Ce_{0.9}Cu_2O_{10-\delta}$ a maximum in –MR is evidenced; at the same time a sharp decline of the maximum magnetisation $M_{max}$ at $T_{Cu}$ is observed, which is consistent with the loss of magnetopolarons in the $CuO_2$ planes. This suggests that both the spin-polarised antiferromagnetic and superconducting precursor phases are present in underdoped ruthenocuprates but the superconducting precursor phase is increasingly stabilized above p = 0.042 so that $-MR_{7T}$(5K) and $M_{max}$ fall rapidly as a result of the onset of d-wave pairing correlations between Cu holes, eventually leading to superconductivity for p > 0.06. Hence the magnetoresistance is sensitive to the competition between antiferromagnetic and superconducting fluctuations in the i-AF phase and could be used to probe the pseudogap state further.

CONCLUSIONS

These results show that the physical properties of the underdoped $RuSr_2(Nd,Y,Ce)_2Cu_2O_{10-\delta}$ ruthenocuprates are dependent on both the $(Nd,Y)^{3+}/Ce^{4+}$ ratio and the oxygen deficiency within the $(Nd,Y,Ce)_2O_{2-\delta}$ block. A full chemical characterisation of samples is therefore important in order to understand the complex physics of the 1222 ruthenocuprates, in which high-Tc superconductivity and large magnetoresistances can emerge from the interactions between weakly ferromagnetic $RuO_2$ layers and doped antiferromagnetic $CuO_2$ planes. Induced antiferromagnetic (1/2 1/2 0) Cu spin order is observed across the pseudogap region up to the onset of superconductivity, and the high field MR varies systematically with Cu hole doping in the $RuSr_2Nd_{1.8-x}Y_{0.2}Ce_xCu_2O_{10-\delta}$ series. A collapse of -MR and magnetisation above p = 0.042 reveals the onset of d-wave spin pairing in the pre-superconducting region. Further research will be needed to investigate whether phase segregation occurs in the pre-superconducting region and to study the low temperature antiferromagnetic to superconducting transition at p > 0.059.

<div align="center">ACKNOWLEDGEMENTS</div>

We thank P. Henry for assistance with the neutron experiments. We also acknowledge the Leverhulme Trust for provision of an early career fellowship (ACM) and financial support, the Ministry of Science and Technology, Government of Pakistan for a studentship (FS), and the UK EPSRC for beam time provision and financial support.

TABLE I.  Variation of hole doping level, $p$, magnetoresistance, $T_{Ru}$, $T_{Cu}$, $M_{max}$ and $\delta$ with $x$ in the RuSr$_2$Nd$_{2.0-x-y}$Ce$_x$Y$_y$Cu$_2$O$_{10-\delta}$ solid solutions; $y = 0.2$ in all samples except where otherwise indicated.

| $p$ | $x$ | Anneal | $\delta$ | MR$_{7T}$ (5 K) (%) | $T_{Ru}$ (K) | $T_{Cu}$ (K) | $M_{max}$ (emu mol$^{-1}$) |
|---|---|---|---|---|---|---|---|
| **0.010** | 0.95 | N$_2$ | 0.015 | -14 | 180 | 110 | 6.57 |
| **0.013** | 0.90 | N$_2$ | 0.037 | -17 | 180 | 115 | 7.98 |
| **0.017** | 0.90 ($y = 0.1$) | air | 0.033 | -29 | 165 | 104 | 5.47 |
| **0.021** | 0.95 | air | 0.004 | -20 | 150 | 77 | 6.00 |
| **0.033** | 0.90 | air | 0.017 | -22 | 130 | 59 | 7.39 |
| **0.042** | 0.85 | air | 0.033 | -23 | 117 | 49 | 4.73 |
| **0.046** | 0.80 | air | 0.054 | -22 | 95 | 41 | 3.02 |
| **0.052** | 0.75 | air | 0.073 | -21 | 85 | 33 | 1.32 |
| **0.055** | 0.80 | O$_2$ | 0.045 | -19 | 90 | 35 | 2.55 |
| **0.055** | 0.70 | air | 0.095 | -19 | 78 | 27 | 0.64 |
| **0.057** | 0.75 | O$_2$ | 0.068 | -17 | 78 | 29 | 0.85 |
| **0.059** | 0.70 | O$_2$ | 0.091 | -15 | 73 | 25 | 0.42 |

TABLE II. Refined atomic parameters for $RuSr_2Nd_{0.85}Ce_{0.95}Y_{0.2}Cu_2O_{10}$ (site occupancies; variable coordinates x,y,z and isotropic thermal parameters U) from variable temperature neutron diffraction data. Atom positions are Ru 2(a) (0, 0, 0), Sr 4(e) (½, ½, z), Nd/Ce/Y 4(e) (½,½, z), Cu 4(e) (0, 0, z), O(1) 16(n) (x, 0, z), O(2) 8(g) (0, ½, z), O(3) 8(j) (x, ½, 0), O(4) 4(d) (0, ½, ¼). Refined cell parameters and selected bond lengths (Å) are also displayed.

| Atom | Occup-ancy | | Temperature (K) | | | | | | | | |
|---|---|---|---|---|---|---|---|---|---|---|---|
| | | | 5 | 30 | 60 | 90 | 120 | 150 | 190 | 220 | 290 |
| Ru | 1.00 | $U_{iso}$ (Å$^2$) | 0.0058(7) | 0.0061(7) | 0.0064(7) | 0.0059(7) | 0.0064(7) | 0.0071(7) | 0.0073(7) | 0.0073(7) | 0.0063(7) |
| Sr | 1.00 | z | 0.07766(9) | 0.07763(9) | 0.07755(9) | 0.07762(9) | 0.07755(9) | 0.07761(9) | 0.07743(9) | 0.07757(9) | 0.07732(9) |
| | | $U_{iso}$ (Å$^2$) | 0.0095(5) | 0.0099(5) | 0.0106(5) | 0.0109(5) | 0.0114(5) | 0.0120(5) | 0.0134(5) | 0.0134(5) | 0.0149(5) |
| Nd/Ce/Y | 1.00 | z | 0.20457(9) | 0.20462(9) | 0.20460(9) | 0.20454(9) | 0.20460(9) | 0.20462(9) | 0.20466(9) | 0.20477(9) | 0.20460(9) |
| | | $U_{iso}$ (Å$^2$) | 0.0037(5) | 0.0038(5) | 0.0040(5) | 0.0039(5) | 0.0051(5) | 0.0044(5) | 0.0054(5) | 0.0055(5) | 0.0065(5) |
| Cu | 1.00 | z | 0.14364(8) | 0.14367(8) | 0.14360(8) | 0.14373(8) | 0.14375(8) | 0.14372(8) | 0.14373(8) | 0.14370(8) | 0.14372(8) |
| | | $U_{iso}$ (Å$^2$) | 0.0037(4) | 0.0038(4) | 0.0036(4) | 0.0039(4) | 0.0039(4) | 0.0044(4) | 0.0046(4) | 0.0046(4) | 0.0053(4) |
| O(1) | 0.25 | x | -0.028(4) | -0.033(4) | -0.036(4) | -0.036(4) | -0.041(4) | -0.041(4) | -0.043(4) | -0.046(4) | -0.046(4) |
| | | z | 0.0668(1) | 0.0668(1) | 0.0667(1) | 0.0667(1) | 0.0666(1) | 0.0665(1) | 0.0665(1) | 0.0663(1) | 0.0662(1) |
| | | $U_{iso}$ (Å$^2$) | 0.016(1) | 0.014(1) | 0.015(1) | 0.014(1) | 0.012(1) | 0.013(1) | 0.013(1) | 0.012(1) | 0.014(1) |
| O(2) | 1.00 | z | 0.15012(6) | 0.15013(6) | 0.15017(6) | 0.15010(6) | 0.15013(6) | 0.15012(6) | 0.15011(6) | 0.15006(6) | 0.15014(6) |
| | | $U_{iso}$ (Å$^2$) | 0.0043(4) | 0.0040(4) | 0.0045(4) | 0.0048(4) | 0.0049(4) | 0.0056(4) | 0.0057(4) | 0.0061(4) | 0.0068(4) |
| O(3) | 0.50 | x | 0.1300(8) | 0.1309(8) | 0.1304(8) | 0.1307(8) | 0.1299(8) | 0.1298(8) | 0.1292(8) | 0.1299(8) | 0.1272(8) |
| | | $U_{iso}$ (Å$^2$) | 0.015(1) | 0.015(1) | 0.015(1) | 0.016(1) | 0.016(1) | 0.017(1) | 0.017(1) | 0.018(1) | 0.019(1) |
| O(4) | 1.00 | $U_{iso}$ (Å$^2$) | 0.0062(6) | 0.0057(6) | 0.0060(6) | 0.0070(6) | 0.0065(6) | 0.0066(6) | 0.0073(6) | 0.0074(6) | 0.0079(6) |
| | | a (Å) | 3.84365(4) | 3.84369(3) | 3.84404(3) | 3.84462(3) | 3.84541(3) | 3.84636(3) | 3.84771(3) | 3.84878(3) | 3.85147(3) |
| | | c (Å) | 28.4934(6) | 28.4932(6) | 28.4945(5) | 28.4971(6) | 28.5017(5) | 28.5069(5) | 28.5170(5) | 28.5256(5) | 28.5480(5) |
| | | V (Å$^3$) | 420.951(10) | 420.959(8) | 421.054(8) | 421.219(8) | 421.461(8) | 421.745(9) | 422.191(9) | 422.553(8) | 423.476(8) |
| | | $\chi^2$ | 1.17 | 1.19 | 1.20 | 1.13 | 1.14 | 1.16 | 1.14 | 1.09 | 0.99 |

| | | | | | | | | | |
|---|---|---|---|---|---|---|---|---|---|
| $R_{WP}$ (%) | 4.24 | 4.27 | 4.30 | 4.17 | 4.19 | 4.21 | 4.17 | 4.08 | 3.82 |
| $R_P$ (%) | 3.09 | 3.19 | 3.18 | 3.07 | 3.12 | 3.15 | 3.12 | 3.04 | 2.79 |
| $d_{RC}$ | 8.185(5) | 8.187(5) | 8.183(5) | 8.192(5) | 8.194(5) | 8.194(6) | 8.198(6) | 8.198(6) | 8.206(6) |
| $d_I$ | 6.062(3) | 6.060(3) | 6.064(3) | 6.057(3) | 6.057(3) | 6.060(3) | 6.061(3) | 6.065(3) | 6.069(3) |
| Cu-O(1) | 2.192(5) | 2.195(5) | 2.195(5) | 2.201(5) | 2.206(5) | 2.207(4) | 2.207(5) | 2.214(5) | 2.220(5) |
| Cu-O(2) | 1.9307(3) | 1.9307(3) | 1.9311(3) | 1.9309(3) | 1.9313(3) | 1.9318(2) | 1.9324(3) | 1.9329(3) | 1.9344(3) |
| Ru-O(1) | 1.906(4) | 1.907(4) | 1.906(4) | 1.905(4) | 1.903(4) | 1.902(4) | 1.905(4) | 1.901(4) | 1.898(4) |
| Ru-O(3) | 1.9858(10) | 1.9866(10) | 1.9863(10) | 1.9869(10) | 1.9866(10) | 1.9869(10) | 1.9870(10) | 1.9882(10) | 1.9871(10) |
| Cu-O(2)-Cu | 169.0(2) | 169.1(2) | 168.9(2) | 169.2(2) | 169.2(2) | 169.2(2) | 169.2(2) | 169.2(2) | 169.1(2) |
| Ru-O(3)-Ru | 150.8(2) | 150.7(2) | 150.8(2) | 150.7(2) | 150.9(2) | 150.9(2) | 151.0(2) | 150.9(2) | 151.4(2) |

TABLE III. Refined cell parameters and selected bond lengths (Å) for the $RuSr_2Nd_{2.0-x-y}Ce_xY_yCu_2O_{10-\delta}$ solid solutions from neutron powder diffraction data recorded at 5 K, y = 0.2 in all samples except where otherwise indicated.

| | | | | | |
|---|---|---|---|---|---|
| $p$ | 0.013 | 0.017 | 0.021 | 0.033 | 0.046 |
| x | 0.90 (N$_2$) | 0.90, y =0.1 | 0.95 | 0.90 | 0.80 |
| $\delta$ | 0.037 | 0.033 | 0.004 | 0.017 | 0.054 |
| $a$ (Å) | 3.84519(3) | 3.84742(4) | 3.84365(4) | 3.84304(4) | 3.84561(4) |
| $c$ (Å) | 28.4906(5) | 28.4936(5) | 28.4934(6) | 28.4881(5) | 28.4821(6) |
| V (Å$^3$) | 421.247(8) | 421.782(9) | 420.951(10) | 420.740(10) | 421.215(9) |
| $\chi^2$ | 5.7 | 1.41 | 1.17 | 3.30 | 1.00 |
| R$_{WP}$ (%) | 1.67 | 4.56 | 4.24 | 4.23 | 3.99 |
| R$_P$ (%) | 3.44 | 3.38 | 3.09 | 3.37 | 3.23 |
| d$_{RC}$ | 8.162(2) | 8.167(5) | 8.185(5) | 8.168(5) | 8.176(5) |
| d$_l$ | 6.083(2) | 6.080(3) | 6.062(3) | 6.076(3) | 6.065(3) |
| Cu-O(1) | 2.170(3) | 2.194(4) | 2.192(5) | 2.173(4) | 2.191(5) |
| Cu-O(2) | 1.9307(2) | 1.9326(3) | 1.9307(3) | 1.9302(3) | 1.9302(3) |
| Ru-O(1) | 1.915(2) | 1.905(3) | 1.906(4) | 1.918(3) | 1.910(3) |
| Ru-O(3) | 1.9815(6) | 1.9876(8) | 1.9858(10) | 1.9827(8) | 1.9820(10) |
| R-O(2) | 2.475(2) | 2.476(2) | 2.470(2) | 2.472(2) | 2.484(2) |
| R-O(4) | 2.325(2) | 2.320(2) | 2.317(2) | 2.320(2) | 2.317(2) |
| Cu-O(2)-Cu | 169.5(1) | 169.0(2) | 169.0(2) | 169.1(2) | 170.0(3) |
| Ru-O(3)-Ru | 152.0(3) | 150.9(2) | 150.8(2) | 151.5(2) | 151.9(2) |

Figure Captions

FIG. 1    Crystal and magnetic structures of $RuSr_2Nd_{1.8-x}Ce_xY_{0.2}Cu_2O_{10-\delta}$. The interplanar separation of $CuO_2$ planes, $d_I$ and thickness of the $CuO_2.SrO.RuO_2.SrO.CuO_2$ ruthenium copper oxide slabs, $d_{RC}$ are labelled.

FIG. 2.    Variable temperature magnetization data (zero-field and field cooled) for the $RuSr_2Nd_{1.8-x}Ce_xY_{0.2}Cu_2O_{10-\delta}$ solid solutions recorded in H = 100 Oe.

FIG. 3.    (Color online) Electronic phase diagram for 1222 type ruthenocuprates. Ordering temperatures for the Ru spins ($T_{Ru}$, squares) and Cu spins ($T_{Cu}$, circles) at hole dopings $p$ < 0.06 are from $RuSr_2Nd_{1.8-x}Y_{0.2}Ce_xCu_2O_{10-\delta}$ phases and $RuSr_2NdY_{0.1}Ce_{0.9}Cu_2O_{9.967}$ in this study. $T_{Ru}$ (open squares) and superconducting critical temperatures (diamonds) for $RuSr_2Gd_{2-x}Ce_xCu_2O_{10-\delta}$ (x = 0.6- 0.8) samples **Error! Bookmark not defined.** are shown in the $p$ = 0.06-0.07 region.

FIG. 4.    Variation of the resistivity with temperature for representative $RuSr_2(Nd,Y,Ce)_2Cu_2O_{10-\delta}$ samples evidencing non-metallic behaviour.

FIG. 5.    Magnetoresistances $MR_H$ (=$\rho$(H) -$\rho$ (0) /$\rho$ (0)) for sintered $RuSr_2Nd_{1.8-x}Y_{0.2}Ce_xCu_2O_{10-\delta}$ materials at 5 K in magnetic fields up to H = 7 T. Doping values $p$ are shown and data for $p$> 0.044 samples are offset by −10% MR for clarity. The inset shows the low field MR for x = 0.8 (as synthesized and annealed in $O_2$) and x = 0.9 (as synthesized and annealed in $N_2$) which evidence microstructural effects such as domain or grain boundary resistances, as the MR-H curvature is negative for as–prepared samples but positive for $O_2$ and $N_2$-annealed materials.

FIG. 6.    Variations of the −$MR_{7T}$(5 K) magnetoresistance and the maximum value of magnetisation ($M_{max}$) with hole doping $p$ in the $RuSr_2Nd_{1.8-x}Y_{0.2}Ce_xCu_2O_{10-\delta}$ series.

FIG. 7.     Rietveld refinement fit to the 4 K SuperD2B neutron diffraction pattern of 1222 type RuSr$_2$Nd$_{0.85}$Y$_{0.2}$Ce$_{0.95}$Cu$_2$O$_{10-\delta}$. Lower, middle and upper reflection markers correspond to RuSr$_2$Nd$_{0.85}$Y$_{0.2}$Ce$_{0.95}$Cu$_2$O$_{10-\delta}$, a trace of a secondary RuSr$_2$(Nd,Y)Cu$_2$O$_8$ phase, and magnetic 1222 phase reflections respectively.

FIG 8      Variations of the 290 K R-O(2) bond-length and the change in slope of d$_I$ at T$_{Ru}$ with $\delta$ in the RuSr$_2$Nd$_{2-x-y}$Y$_y$Ce$_x$Cu$_2$O$_{10-\delta}$ solid solutions.

FIG. 9.     Temperature variations of (a) unit-cell volume and (b) lattice parameters $c$ and $a$ (inset) for representative RuSr$_2$(Nd,Y,Ce)$_2$Cu$_2$O$_{10-\delta}$ samples.

FIG. 10.    Temperature variations of the interplanar CuO$_2$ separation, d$_I$, and the thickness of the ruthenocuprate slabs, d$_{RC}$, (as defined on Fig. 1) in (a) RuSr$_2$Nd$_{1.0}$Y$_{0.1}$Ce$_{0.9}$Cu$_2$O$_{10-\delta}$ (p = 0.017, $\delta$ = 0.033), (b) RuSr$_2$Nd$_{0.85}$Y$_{0.2}$Ce$_{0.95}$Cu$_2$O$_{10-\delta}$ (p = 0.021, $\delta$ = 0.004), (c) RuSr$_2$Nd$_{0.9}$Y$_{0.2}$Ce$_{0.9}$Cu$_2$O$_{10-\delta}$ (p = 0.033, $\delta$ = 0.017) and (d) RuSr$_2$Nd$_{1.0}$Y$_{0.2}$Ce$_{0.8}$Cu$_2$O$_{10-\delta}$ (p = 0.046, $\delta$ = 0.054).

FIG. 11.    Temperature variations of the integrated (1/2 1/2 2) (bold symbols) and composite (1/2 1/2 0)+(1/2 1/2 1/2) (open symbols) magnetic diffraction peak intensities (relative to 4 K values) for RuSr$_2$NdY$_{0.1}$Ce$_{0.9}$Cu$_2$O$_{9.967}$ (p = 0.017), RuSr$_2$Nd$_{0.9}$Y$_{0.2}$Ce$_{0.9}$Cu$_2$O$_{9.983}$ (p = 0.033) and RuSr$_2$Nd$_{1.0}$Y$_{0.2}$Ce$_{0.8}$Cu$_2$O$_{9.955}$ (p = 0.055) showing the Ru and Cu spin ordering transitions. Critical law fits in the range T$_c$/2 < T < T$_c$ are shown except for the p =0.033 and p = 0.055, I(1/2 1/2 2) data which are fitted by an arbitrary curve.

Fig. 1.

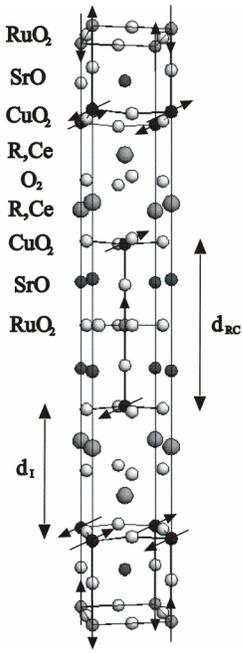

Fig. 2

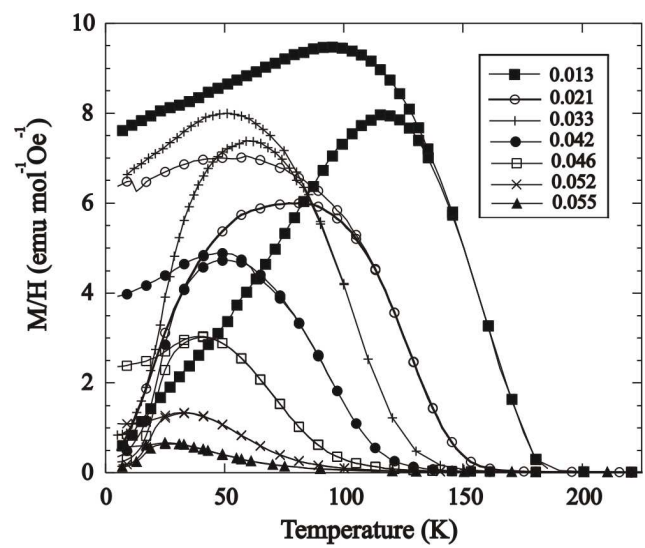

Fig. 3

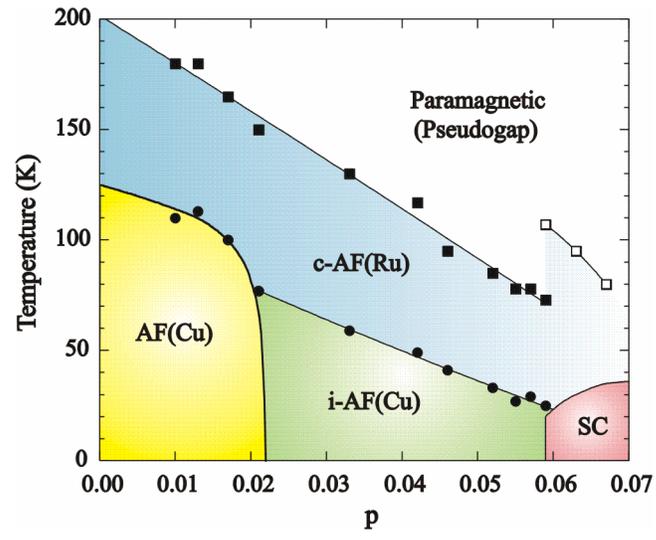

Fig. 4.

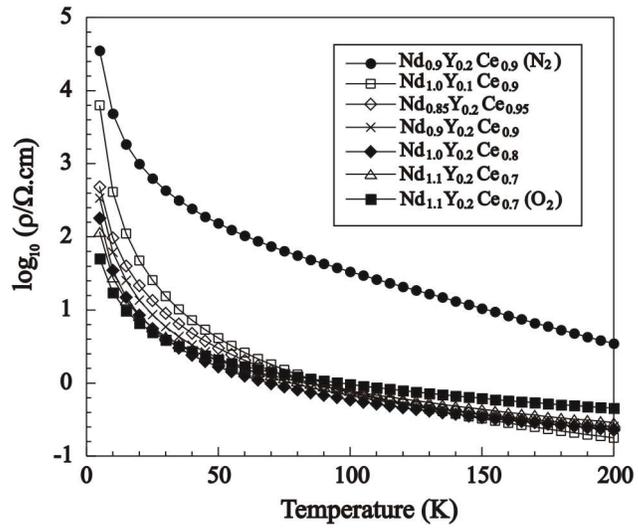

Fig. 5.

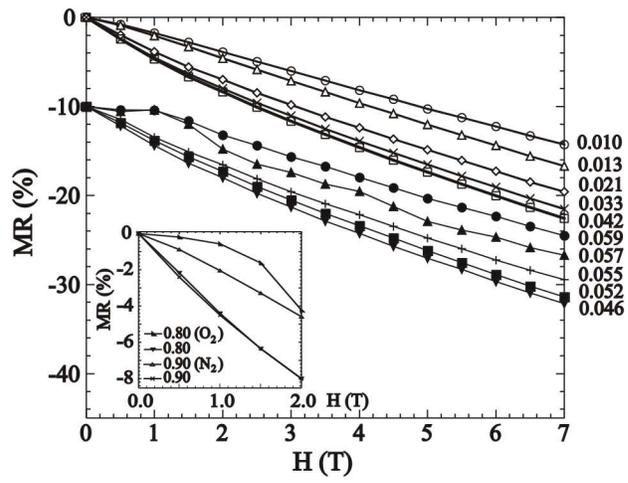

Fig. 6.

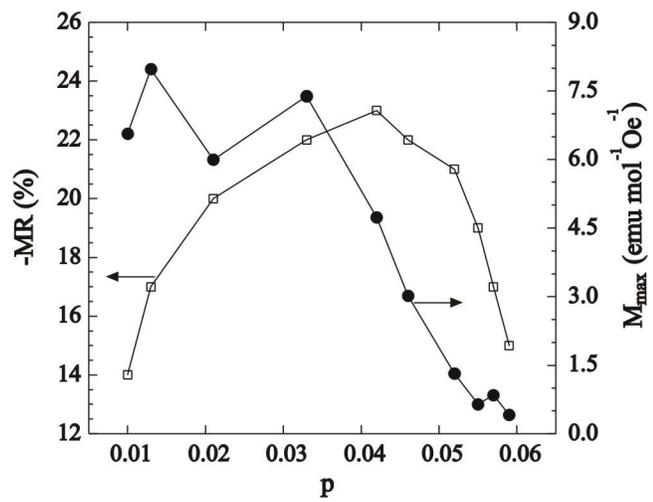

Fig. 7.

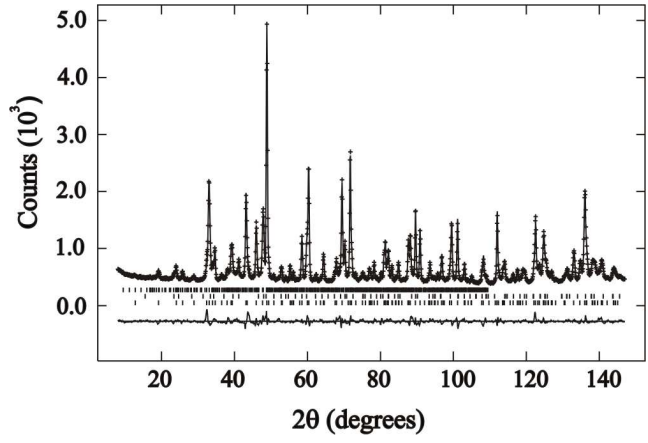

Fig. 8.

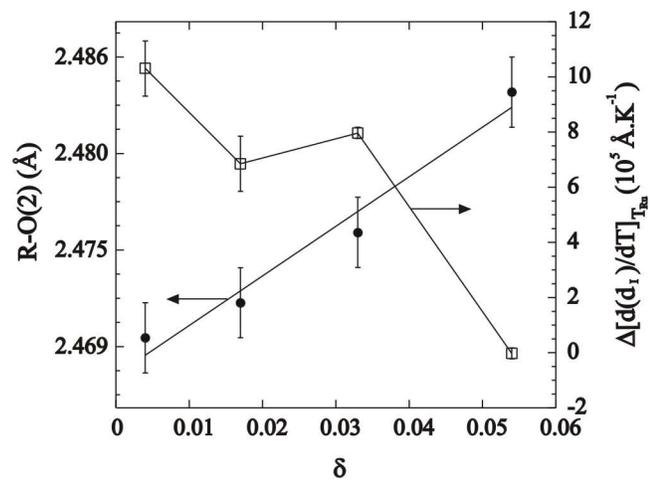

Fig. 9.

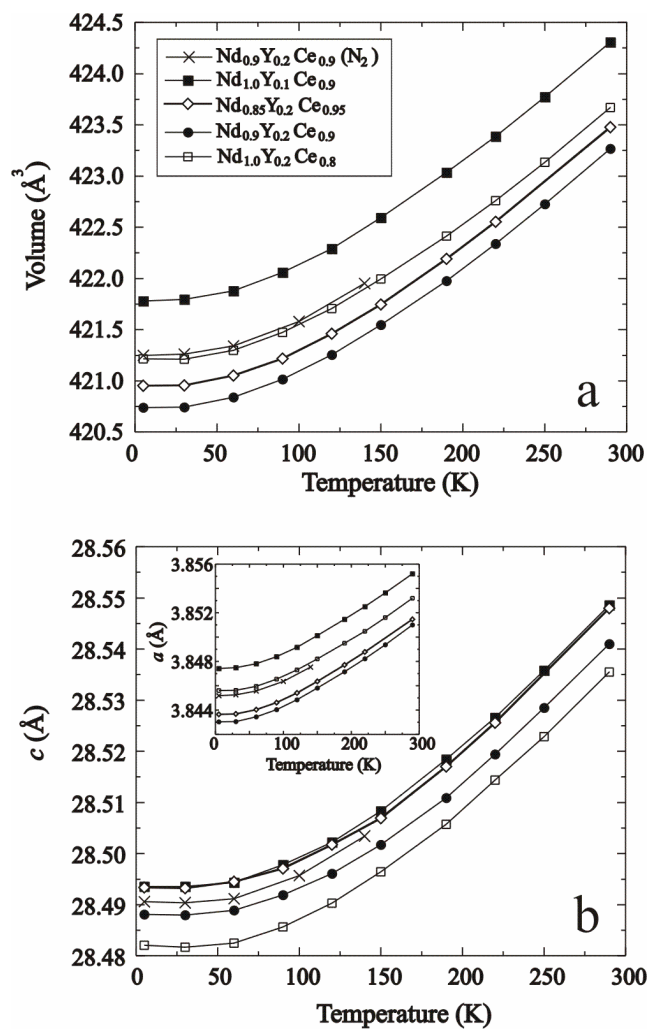

Fig. 10.

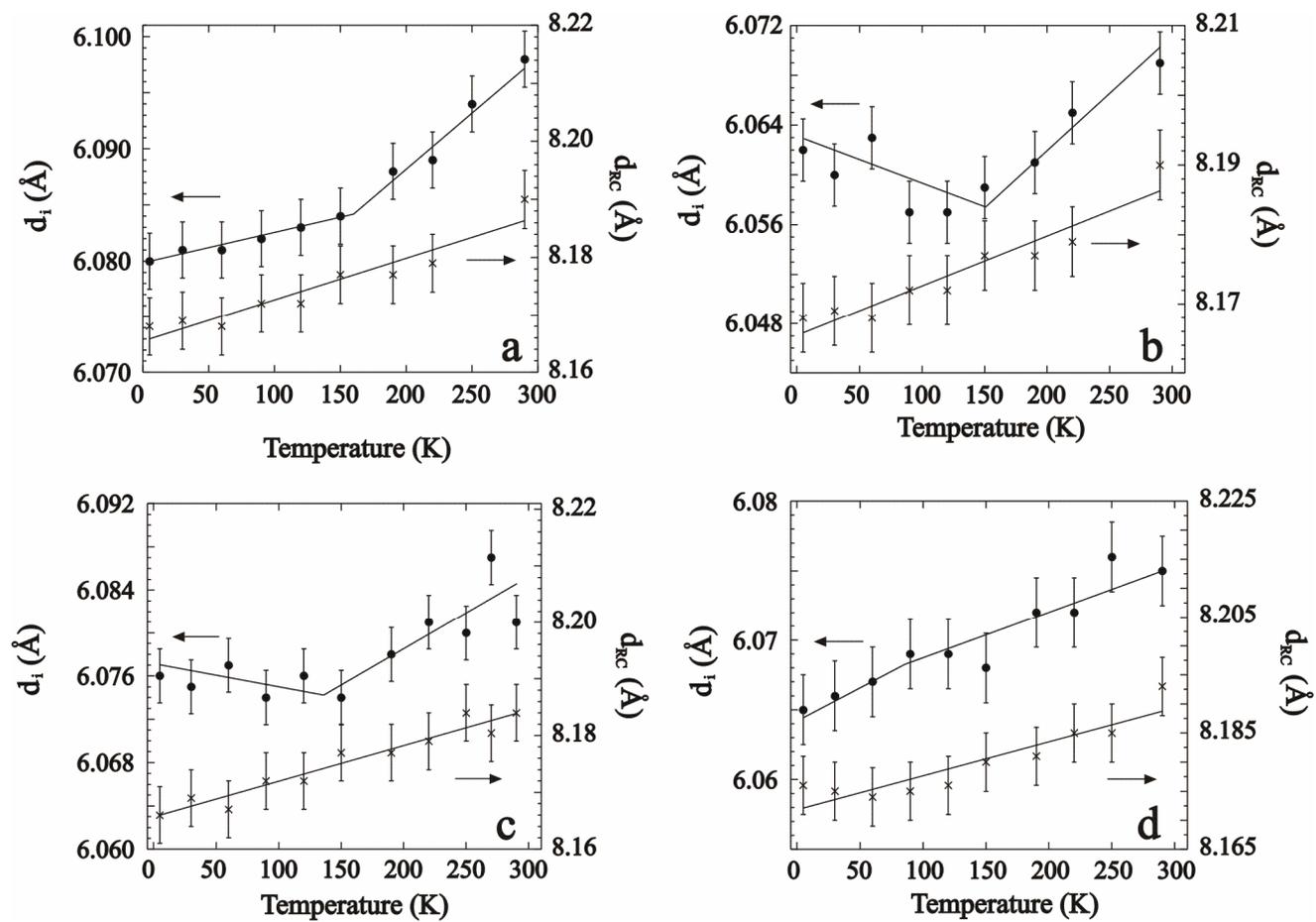

Fig. 11.

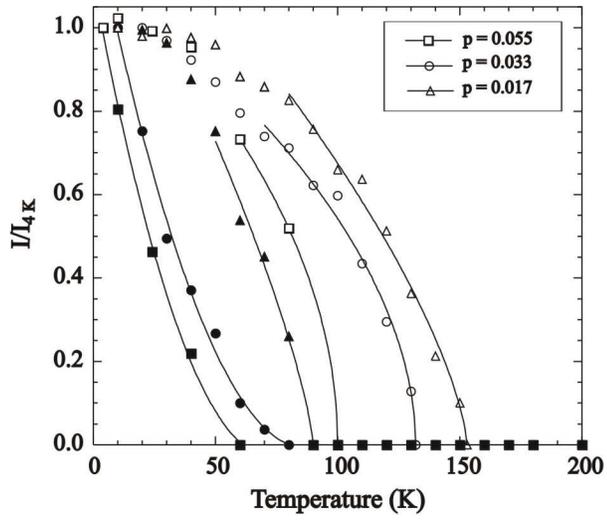